\documentclass[epjCONF]{svjour}
\usepackage{graphics}
\usepackage[varg]{txfonts} 
\usepackage[latin1]{inputenc}
\session-title{Conference Title, to be filled}
\begin{document}

\def\deg{\hbox{$\null^\circ$}}                                                                                    
\def\Msun{\,\textrm {M}_\odot}                                              
\def\kpc{\,{\rm kpc}}                                                           
\def\Gyr{\,{\rm Gyr}}                                                           
\def\Myr{\,{\rm Myr}}                                                           
\def\kms{\,{\rm km}\,{\rm s}^{-1}}
\def\ud{\mathrm{d}}

\title{Could the Hercules satellite be a stellar stream in the Milky Way halo?}
\author{Shoko Jin\inst{1}\fnmsep\thanks{Alexander von Humboldt research fellow}\thanks{\email{shoko@ari.uni-heidelberg.de}} \and Nicolas F. Martin\inst{2}}
\institute{Astronomisches Rechen-Institut, Zentrum f\"ur Astronomie der Universit\"at Heidelberg, M\"onchhofstr. 12--14, D-69120 Heidelberg, Germany \and Max-Planck-Institut f\"ur Astronomie, K\"onigstuhl 17, D-69117 Heidelberg, Germany}
\abstract{
We investigate the possibility that Hercules, a recently discovered Milky Way (MW) satellite, is a stellar stream in the process of formation.  This hypothesis is motivated by Hercules' highly elongated shape as well as the measurement of a tentative radial velocity gradient along its body \cite{aden09vel}. The application of simple analytical techniques \cite{jin07} on radial velocity data of its member stars provides tight constraints on the tangential velocity of the system ($v_\mathrm{t} = -16^{+6}_{-22}\kms$, relative to the Galactic Standard of Rest).  Combining this with its large receding velocity ($145\kms$) and distance ($138\kpc$) yields an orbit that would have taken Hercules to within $6^{+9}_{-2}\kpc$ of the Galactic centre approximately $0.6\Gyr$ ago.  This very small perigalacticon can naturally explain the violent tidal destruction of the dwarf galaxy in the MW's gravitational potential, inducing its transformation into a stellar stream. 
} 

\maketitle
\section{Introduction}
\label{intro}

The Hercules stellar system is one of the so-called `ultra-faint' satellites of the Milky Way.  Residing at a heliocentric distance of $138\kpc$, its highly elongated morphology (ellipticity of 0.65, \cite{coleman07,martin08,sand09}) begs the question of how it could have become so distorted whilst currently inhabiting the outer domains of the stellar halo.  The detection of a tentative radial velocity gradient in its member stars \cite{aden09vel} adds further enigma to this $M_V = -6.6$ \cite{martin08} system.  In this work \cite{hercules10}, we investigate whether Hercules could have undergone a relatively recent close encounter with the Galaxy, resulting in its morphological distortion from a regular dwarf spheroidal galaxy into the highly elongated structure observed today.

\section{Determining the orbit for Hercules}
\label{sec:orbit}

Ascertaining the proximity of Hercules' last perigalactic passage requires its orbit to be calculated.  In the following sections, we first provide a brief description of how the tangential velocity of an object may be determined from a radial velocity gradient measured along it.  This is followed by an outline of the model used to deduce the velocity distribution of Hercules stars and the results of our calculations.  In this study, we use a model for the Milky Way composed of a Miyamoto-Nagai disk and bulge \cite{miyanagai75}, and an NFW \cite{nfw96} profile with parameters constrained by Xue et al. \cite{xue08}.

\subsection{Finding the tangential velocity}
\label{subsec}

For a given Galactic potential $\psi$, the radial velocity gradient along a system can be expressed in terms of its heliocentric distance, $D$, radial velocity, $v_\mathrm{r}$, and tangential velocity, $v_\mathrm{t}$, as follows \cite{jin07}:\footnote{Unless stated otherwise, velocities are quoted relative to the Galactic Standard of Rest and distances are given as heliocentric quantities.}
\begin{equation}
\label{eqn:radvelrun}
\frac{\ud v_\mathrm{r}}{\ud\chi} = v_\mathrm{t} + \left(\nabla_\mathrm{r}\psi\right)\frac{D}{v_\mathrm{t}}\;,
\end{equation}
where $\chi$ is the angle along the stream and $\nabla_\mathrm{r}\psi$ is the component of the gravitational acceleration due to the Galactic potential along the line of sight to the system.  This results in the following two solutions for the tangential velocity of the orbit at the location of the system:
\begin{equation}
\label{eqn:vt}
v_\mathrm{t} = \frac{1}{2} \left( \frac{\ud v_\mathrm{r}}{\ud\chi} \pm \sqrt{\left(\frac{\ud v_\mathrm{r}}{\ud\chi}\right)^2 - 4\left(\nabla_\mathrm{r}\psi\right)D}\right)\;.
\end{equation}
For Hercules, all variables and parameters on the right-hand side of this equation are either known or calculable, hence the direct measurement of a velocity gradient over the body of the satellite translates to only two possible orbits, corresponding to the positive and negative solutions of $v_\mathrm{t}$.

\subsection{Model and results}
\label{subsec:results}

\begin{figure}
\begin{center}
\resizebox{1.0\columnwidth}{!}{\includegraphics{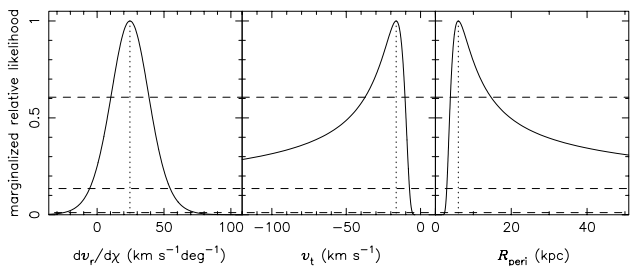}}
\caption{Marginalized relative likelihood distributions for the radial velocity gradient, tangential velocity and perigalacticon, whose best values are $\mathrm{d}v_\mathrm{r}/\mathrm{d}\chi = 25\pm15\kms/\mathrm{deg}$, $v_\mathrm{t} = -16^{+6}_{-22}\kms$ and $R_\mathrm{peri} = 6^{+9}_{-2}\kpc$, respectively.  From top to bottom, the horizontal, dashed lines intersect the likelihood distributions at the boundaries of the 1-, 2- and 3-$\sigma$ confidence intervals.
\label{fig:likelihood}}
\end{center}
\end{figure}

We employ a maximum likelihood algorithm on Hercules member stars with radial velocity determinations \cite{aden09vel} to find the likelihood distribution functions of the following properties of the stellar system: the mean radial velocity of the member stars, $\bar{v}_\mathrm{r}$, the radial velocity gradient, $\ud v_\mathrm{r}/\ud\chi$, and the velocity dispersion of the stars, $s$, around the velocity gradient.  
Under our assumption that Hercules is a stellar stream, the direction of its motion on the sky is given by the direction of its elongation, in other words the major axis of the system.  The position angle of the satellite \cite{martin08} is therefore used as a prior in deducing the likelihood distributions, and hence the best model.

\begin{figure}
\begin{center}
\resizebox{0.75\columnwidth}{!}{\includegraphics{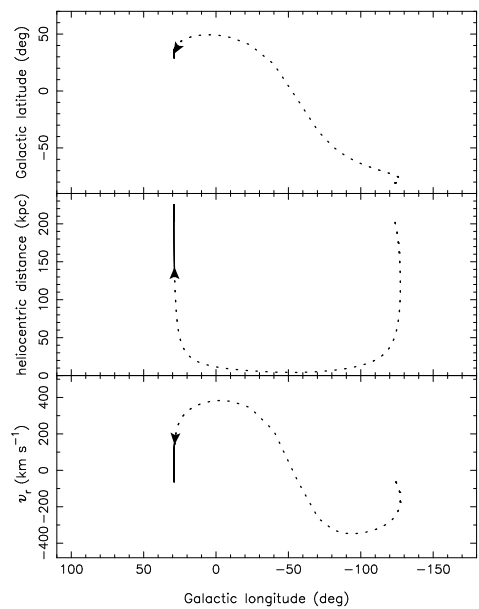}}
\caption{\label{lbdv}Orbit of the Hercules stellar stream using our best model as given in the text, integrated forward and backward from the present location for $2\Gyr$ in each direction.  The panels show the sky projection of the orbit in Galactic coordinates (top), its heliocentric distance (middle) and radial velocity relative to the Galactic Standard of Rest (bottom).  The solid and dotted lines respectively denote the forward and backward-integrated orbit from the present location of Hercules, which is indicated by an arrowhead in the direction of the orbit.\label{fig:orbit}}
\end{center}
\end{figure}

Using equation (\ref{eqn:vt}), the likelihood distribution function of the radial velocity gradient can be transformed into that for the tangential velocity.  The distance is assumed to be $D = 138\pm7\kpc$, obtained by averaging the distance measurements of Ad\'en et al. \cite{aden09dist} and Sand et al. \cite{sand09}.  Each value of the tangential velocity completes the 6D phase-space information necessary to derive an orbit and therefore translates to a single pericentric distance.  This leads to the marginalized relative likelihood distributions shown in Figure~\ref{fig:likelihood} for the radial velocity gradient, tangential velocity and perigalacticon, with the best values and 1-$\sigma$ errors\footnote{This corresponds to the 1-$\sigma$ confidence interval of the likelihood distribution, defined such that the relative likelihood drops by a factor $\mathrm{e}^{-k^2/2}$ at the boundaries of a $k$-$\sigma$  confidence-interval range.} for these parameters being $\ud v_\mathrm{r}/\ud\chi = 25\pm15\kms/\mathrm{deg}$, $v_\mathrm{t} = -16^{+6}_{-22}\kms$ and $R_\mathrm{peri} = 6^{+9}_{-2}\kpc$, respectively.  These constitute our best model for the Hercules system under the assumptions and prior stated above, and the properties of the corresponding orbit are shown in Figure~\ref{fig:orbit}.  Employing the other possible solution for the tangential velocity from equation (\ref{eqn:vt}) leads to orbits that never venture close enough to the Galactic centre to have been affected by tides, and are therefore considered not relevant for the purpose of this study.

\section{Summary}
\label{sec:summary}

The technique for determining the tangential motion of a stream via its radial velocity gradient is a powerful tool, whose strength in aiding orbit determinations has a particularly timely role in the era of large sky photometric and spectroscopic surveys.
We use this technique and a maximum likelihood algorithm to find a viable orbit for the Hercules satellite that would have taken the stellar system to within $6\kpc$ of the Galactic centre $\sim0.6\Gyr$ ago.  Such a small perigalacticon could explain a disruption of the system so violent that we now observe it as an unbound stellar stream.  This simple possibility for Hercules should serve as a cautionary warning for treating all recently discovered ultra-faint dwarf galaxies as a single population of bound stellar systems.

\end{document}